\documentclass[twocolumn]{aastex631}
\AtBeginDocument{}
\usepackage{amsmath,amssymb}
\graphicspath{{./}}
\newcommand{\Mej}{M_{\rm ej}}
\providecommand{\Msun}{}
\renewcommand{\Msun}{\,M_\odot}
\shorttitle{Terminal instability of the Solar System}
\shortauthors{Batygin et al.}

\begin{document}

\title{Terminal instability of the Solar System triggered by stochastic solar mass loss}

\author{Konstantin Batygin}
\affiliation{Division of Geological and Planetary Sciences, California Institute of Technology, Pasadena, CA 91125, USA}

\author{Jim Fuller}
\affiliation{TAPIR, Walter Burke Institute for Theoretical Physics, California Institute of Technology, Pasadena, CA 91125, USA}

\author{Fred C. Adams}
\affiliation{Department of Physics, University of Michigan, Ann Arbor, MI 48109, USA}
\affiliation{Department of Astronomy, University of Michigan, Ann Arbor, MI 48109, USA}

\begin{abstract}
From its birth, celestial mechanics has been deeply intertwined with the question of the Solar System's dynamical stability. For the inner planets, this question is now statistically settled: Mercury's orbit carries a $\sim$1\% chance of destabilization before the Sun leaves the main sequence. The outer Solar System has seemed more secure, with its intrinsic dynamical lifetime estimated at $\sim$$10^{18}$ years. Even accounting for the Sun's mass loss and stellar flybys, the orbital architecture of the giant planets had been expected to persist for $\sim$100\,Gyr. Here we show that these estimates rest on the assumption that solar mass loss is smooth. The recently measured white dwarf recoil demands asymmetric mass loss that is most readily attributed to discrete, independently directed ejections that impulsively perturb stellar motion. As the Sun sheds its envelope in such parcels, the planets' orbits random-walk with amplitude set by the mass-loss granularity. At a coarseness corresponding to observationally permitted kicks, this stochastic forcing restructures the outer Solar System concurrently with the Sun's death. In particular, our numerical experiments reveal that orbit crossings can commence on the red giant branch, with $\sim$40\% of realizations undergoing disruption or violent scattering before the white dwarf forms and $\sim$90\% self-destructing within 3\,Gyr. The outer Solar System's dynamical lifetime thus collapses from $10^{18}$ years to approximately a gigayear after white dwarf formation.
\end{abstract}


\section{Introduction}

Whether the planets are stable is the question around which celestial mechanics was built. Newton doubted it, suspecting that the mutual perturbations of Jupiter and Saturn---manifest in the slow drift of their orbits that Kepler had already noticed---would eventually unravel the planetary order without divine intervention \citep{wilson1985,laskar1996}. Laplace and Lagrange answered him at second order in the eccentricities and inclinations, demonstrating that secular perturbations leave the semimajor axes invariant and merely cycle the eccentricities. Moreover, Laplace's resolution of the Jupiter--Saturn ``Great Inequality'' as a periodic, near-resonant modulation stands as the emblematic rescue of Solar System stability \citep{laskar2013}. Poincar\'e subsequently showed, however, that the rescue could never be complete. The modern, computational era has rendered the final verdict statistical: the inner Solar System is marginally, but quantifiably stable, with Mercury's orbit carrying a $\sim$1\% probability of destabilization within the Sun's remaining main-sequence lifetime (with general relativity playing an interesting stabilizing role) \citep{laskar1989,sussmanwisdom1992,batyginlaughlin2008,laskar2009,batygin2015,zeebe2015,woillez2020}.

For the outer Solar System, the question is sharper and the answer has seemed more secure. The four giant planets, weakly chaotic through the interaction of their near-commensurabilities \citep{murrayholman1999,varadi1999,michtchenko2001}, are intrinsically stable on timescales vastly exceeding the age of the Universe. To this end, \citet{murrayholman1999} estimated a timescale of $\sim$$10^{18}$ years for chaotic diffusion to destabilize Uranus. On such horizons, extrinsic astrophysical effects intervene first. The Sun will shed $\sim$46\% of its mass during its giant-branch evolution \citep{sackmann1993,schroder2008}, adiabatically doubling the planets' orbital radii \citep{hadjidemetriou1963,duncanlissauer1998,veras2011}, and the expanded system presents a larger cross-section to passing stars. \citet{zink2020} argued on this basis that stellar flybys would dismantle the giant-planet system within $\sim$30--100\,Gyr. The consensus picture, then, holds that the outer Solar System's demise is remote and externally inflicted.

In this work, we show that this picture fails at its foundation, because solar mass loss will not be smooth. Gaia astrometry of wide binaries reveals that newborn white dwarfs receive velocity kicks of $\sim$0.75\,km\,s$^{-1}$ \citep{elbadryrix2018,hwang2025}, corroborated by the depletion and extended spatial distributions of white dwarfs in star clusters \citep{fellhauer2003,davis2008,richer2021,miller2022}. The emerging explanation is that red-giant envelopes depart in $N \sim M_{\rm env}/\Mej$ discrete, asymmetric ejection events. Physically, these events correspond to plumes launched by pulsation-assisted convection, which 3D simulations predict and which interferometric imaging of evolved stars resolves as discrete, evolving dust clouds \citep{ohnaka2016,freytag2023,fullertsuna2024,ma2025}. Each ejection imparts a small recoil $\sim(\Mej/M_\star)\,v_{\rm esc}$ to the star in a random direction, and these perturbations accumulate through a random walk to the observed kick amplitude \citep{fuller2026}. Critically, each impulsive kick onto the Sun displaces every heliocentric orbit; thousands of them constitute a stochastic forcing that the classical stability analyses---from the invariability theorems of Laplace and Lagrange to the $10^{18}$-year bound or the flyby-erosion estimates---never contemplated. Applied as a single terminal impulse, the observed kick is already known to destabilize wide multi-planet systems \citep{stephan2026}; resolving that impulse into its physical constituents, as we do here, moves the onset of instability into the mass-loss epoch itself and ties the dynamical outcome to the granularity of the mass-ejection events. The consequences, we find, are not a perturbative correction to the old answer but its wholesale replacement. The outer Solar System's dynamical lifetime is set by the granularity of its star's mass loss, and at the parcel masses implied by the episodic interpretation of the observed kicks, instability begins during the epoch of mass loss itself.

\section{Numerical Experiments}
\label{sec:numerics}

The numerical experiments carried out in this work employ standard community tools---the \textsc{rebound} $N$-body package \citep{reinliu2012} with the IAS15 \citep{reinspiegel2015} and TRACE \citep{lu2024} integrators, together with stellar-evolution tracks computed with MESA \citep{paxton2019}---implemented through conventional computational procedures. The complete numerical setup and the implementation of stochastic mass loss are outlined in Appendix~\ref{app:setup}; classification criteria and various robustness tests are collected in Appendix~\ref{app:validation}.

\subsection{Smooth mass loss preserves the Solar System}

We first establish the baseline. We integrate the four giant planets, initialized from their present state vectors, through the Sun's post-main-sequence mass loss (Appendix~\ref{app:setup}), treating $M_\star(t)$ at three levels of realism: linear decay over $10^3$--$10^6$\,yr, two-stage profiles ending in a fast superwind, and the full track of a MESA \citep{paxton2019} solar model evolved from the main sequence to a $0.538\Msun$ white dwarf. In every adiabatic realization ($>$60 integrations spanning these profiles and all secular phases), the outcome is qualitatively identical: semimajor axes expand by the factor $M_{\star,0}/M_{\star,f} \simeq 1.85$, all period ratios remain frozen (Jupiter--Saturn at 2.481 in our de-aliased mean-motion measure), the secular structure---the amplitudes and phases of the eccentricity and inclination eigenmodes---is preserved, and the resonant angles of every nearby commensurability circulate throughout. The post-main-sequence Solar System, under smooth mass loss, is a uniformly enlarged copy of the present one\footnote{In smooth-mass-loss simulations, \citet{zink2020} reported capture of Jupiter and Saturn into the 5:2 mean-motion resonance. In revisiting that work, we find that this capture cannot be reproduced and is consistent with a coordinate-handling artifact.}, and correspondingly stable \citep{duncanlissauer1998}. The expanded control system thus survives multi-gigayear integrations without incident.

\subsection{Episodic mass loss rearranges the Solar System}

Everything changes when the envelope departs in pieces. Following the recently proposed episodic-ejection framework \citep{fuller2026}, we model the mass loss as $N = \Delta M/\Mej$ discrete events distributed along the MESA track in proportion to the local mass-loss rate, each removing $\Mej$ from the Sun and applying an impulsive, isotropically oriented recoil $\Delta v = (\Mej/M_\star)\,v_{\rm esc}(t)$ (Appendix~\ref{app:setup}). Discreteness is essential: smooth-but-anisotropic winds average away at planetary distances \citep{veras2013aniso}, but impulsive recoils, uncorrelated event to event, evade secular averaging entirely---the same reason stochastic forcing restructures resonant dynamics in turbulent disks \citep{adams2008,reinpap2009,paardekooper2013,batyginadams2017} and in Neptune's planetesimal-driven migration \citep{nesvorny2016}. In the present case, the noise source is the star itself, and we remark that both temporal discreteness and per-event asymmetry are strictly necessary for large-scale excitation (in control simulations with stochastically timed but spherically symmetric ejection events, the evolution reduces to the adiabatic case and remains stable for 3\,Gyr; Appendix~\ref{app:setup}). 

For the fiducial $\Mej \sim 10^{-4}\Msun$ case suggested by ejection models and consistent with the observed kick amplitudes \citep{elbadryrix2018,fuller2026}, the Sun receives $N \simeq 4{,}600$ recoils of several m\,s$^{-1}$ (median $\sim$7\,m\,s$^{-1}$). Individually these are minuscule; collectively they are transformative. Because successive recoils strike at uncorrelated orbital phases, every orbital element random-walks, with accumulated period-ratio dispersion $\sigma \propto \sqrt{N}\,\Mej \propto \sqrt{\Mej}$---a scaling our 672-realization grid confirms across three decades (Fig.~1). At fiducial $\Mej$, the dispersion of the final period ratio of Saturn and Jupiter (unperturbed value 2.481) reaches $\sigma = 0.23$ when ejections are restricted to the asymptotic giant branch (AGB) and $\simeq$0.6 when they operate throughout the giant-branch phases. This is an order of magnitude beyond anything in the adiabatic picture. Meanwhile, Neptune's eccentricity is pumped as high as $\sim$0.5, and mutual inclinations diffuse to several degrees. The neat concentric architecture that every prior stability analysis took as its initial condition is replaced by a broad distribution of eccentric, reordered, and compressed configurations (Fig.~2).

\begin{figure}[!tp]\centering
\includegraphics[width=\linewidth]{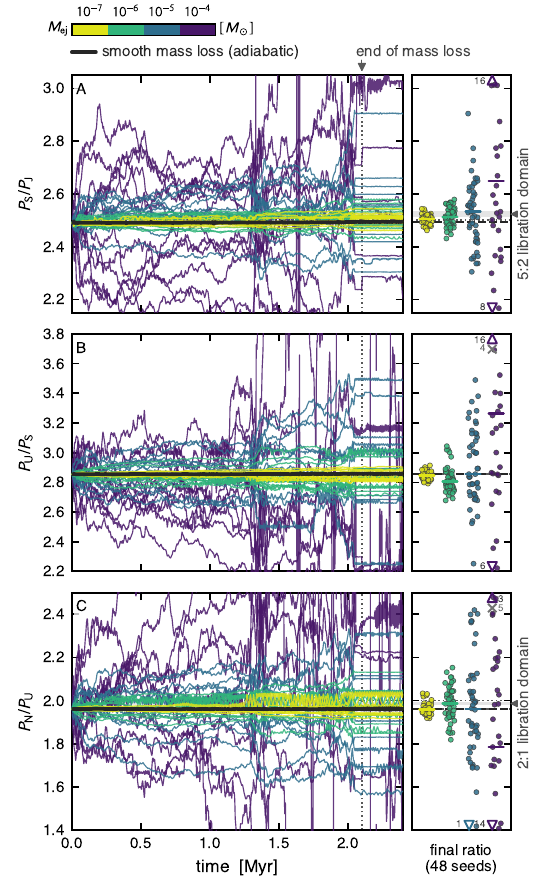}
\caption{\textbf{Episodic mass loss randomizes the giant planets' period ratios.} Evolution of $P_{\rm S}/P_{\rm J}$, $P_{\rm U}/P_{\rm S}$, and $P_{\rm N}/P_{\rm U}$ through the mass-loss epoch for the four decade-spaced ejection masses $\Mej = 10^{-7}$, $10^{-6}$, $10^{-5}$, and $10^{-4}\Msun$ (color scale; twelve realizations each, all-phases variant), against the adiabatic track (black), which conserves all three ratios. Right margins: final-ratio distributions (48 realizations per $\Mej$); horizontal bars mark cell medians, count-bearing triangles tally seeds beyond the plotted range, and gray crosses count realizations whose ratio is undefined after an ejection leaves an orbit unbound. Gray bands: the 5:2 libration domain in panel A and the 2:1 domain in panel C, mapped at the final stellar mass $0.54\Msun$ at adiabatic eccentricities; dotted lines mark the commensurabilities. Across this range of $\Mej$ the accumulated stellar kick speed $\mathcal{D}_v^{1/2}$ spans 0.018--0.58\,km\,s$^{-1}$.}
\end{figure}

\begin{figure*}[!tp]\centering
\includegraphics[width=\textwidth]{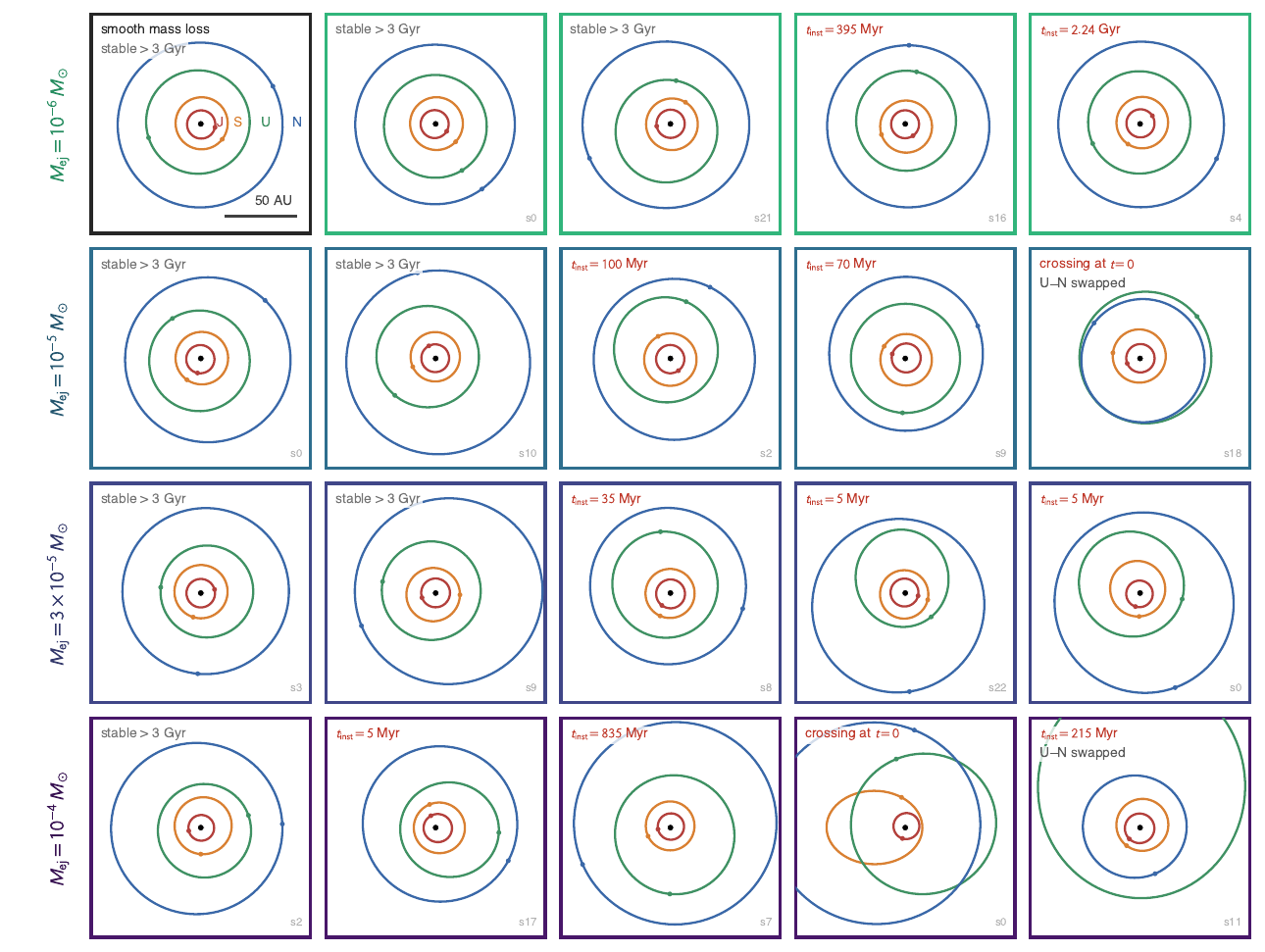}
\caption{\textbf{Examples of post-mass-loss orbital architectures.} Orbit plots of representative end states (rows: increasing $\Mej$; top-left: the adiabatic control), each annotated with its subsequent fate from 3-Gyr integrations; instability times are measured from white dwarf formation, so $t=0$ marks the end of mass loss. Visually pristine systems can carry Gyr-delayed instabilities; at fiducial $\Mej$, crossing and swapped configurations exist already at the moment of white dwarf formation.}
\end{figure*}

\subsection{Instability begins on the red giant branch}

The central consequence of stochastic forcing is temporal. The onset of instability, previously consigned to timescales of $10^{18}$ years (intrinsic) or $10^{2}$ gigayears (flyby-assisted), erupts \emph{during the Sun's giant-branch evolution}. In the fiducial all-phases variant of our simulations (which extends episodic ejection beyond the AGB, where it is best calibrated), orbit-crossing configurations first arise before the mass loss is complete in 37 of 48 realizations (77\%; $\Mej = 10^{-4}\Msun$), with the earliest crossings occurring while the Sun is still a red giant of $0.89\Msun$. The instability of the outer Solar System, in these realizations, is not merely hastened by stellar death but is contemporaneous with it. By the time the white dwarf forms, the crossings have progressed to outright disruption or violent scattering in 40\% of fiducial all-phases realizations (19/48). Simulated instances include runs where Saturn is ejected within a few Myr, Uranus and Neptune exchange order, and perihelia are driven inside Jupiter's orbit (Fig.~3).

\begin{figure*}[!tp]\centering
\includegraphics[width=\textwidth]{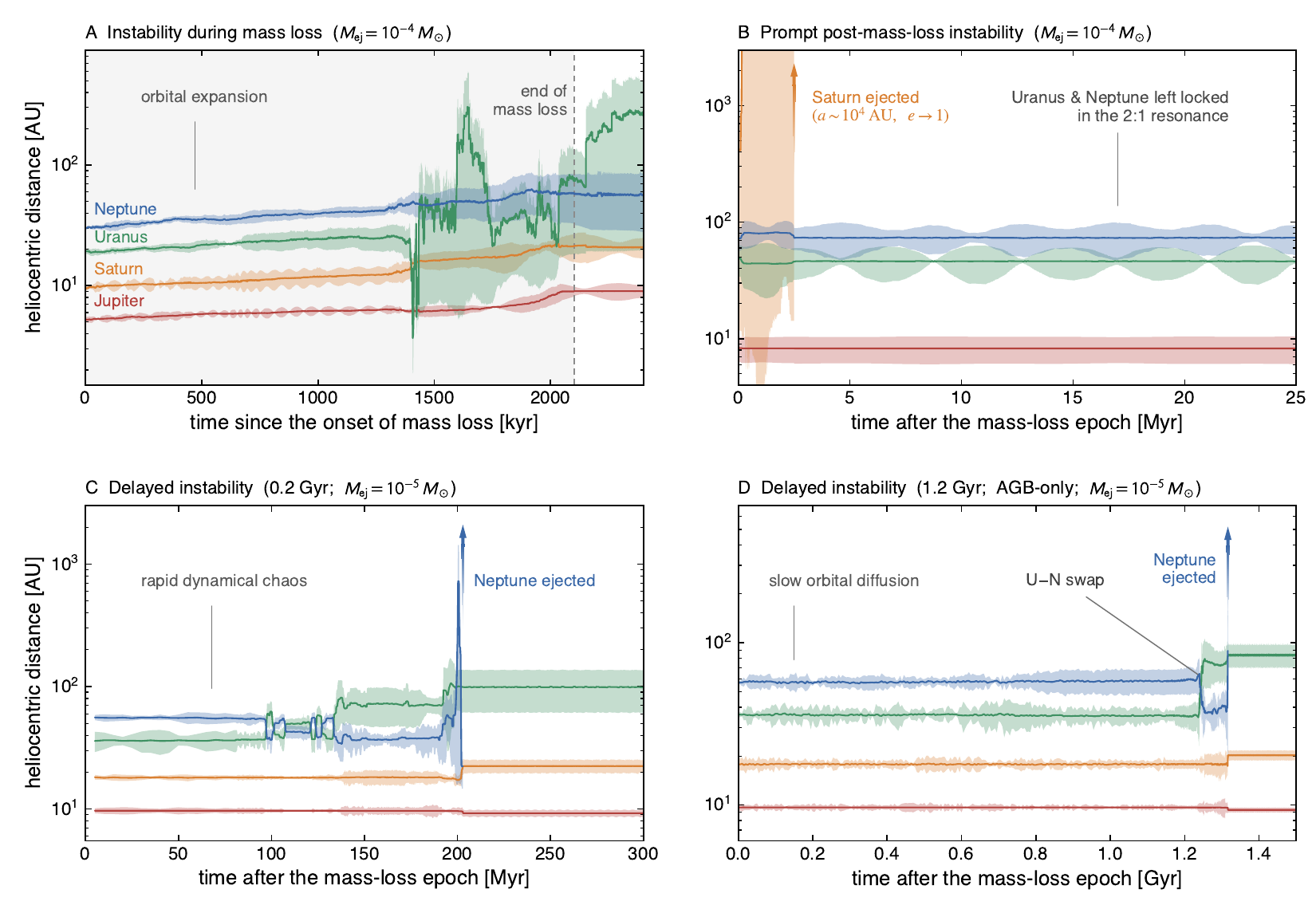}
\caption{\textbf{Instabilities unfolding.} Semimajor axes and perihelion--aphelion envelopes for four representative disintegrations: during the mass-loss epoch (orbit crossing while the Sun is still losing mass); promptly after white dwarf formation (Saturn scattered and ejected within 2.5\,Myr, leaving Uranus and Neptune locked in a phase-protected 2:1 resonance); delayed by 0.2\,Gyr (Uranus--Neptune crossings from $\sim$0.1\,Gyr; Neptune ejected at 0.2\,Gyr); and delayed by 1.2\,Gyr (Uranus--Neptune exchange at 1.24\,Gyr, Neptune ejected at 1.32\,Gyr).}
\end{figure*}

The AGB-only variant of our simulations, kick-free by construction through the red giant branch, produces no red-giant-branch disruption at all, yet converges to the same endpoint: when all 672 end states are integrated forward 3\,Gyr with no external perturbations (Appendix~\ref{app:validation}), the intrinsic instability fraction climbs from $\lesssim$2\% at $\Mej \le 3\times10^{-7}\Msun$ to 96\% (all phases) and 83\% (AGB-only) at fiducial $\Mej$, with onset times spread from a few Myr to beyond 2.5\,Gyr, broadly distributed in $\log t$ (Fig.~4). The vulnerability is legible in the architecture, with instability times cleanly organized by the minimum adjacent orbital separation in mutual Hill units: survival rises from $\sim$3\% below 3, through a broad transition at 3--5, to $\gtrsim$85\% above 5 (the unkicked system sits at 5.5). Passing stars, the linchpin of the previous account \citep{liadams2015,zink2020,kaibraymond2025}, are simply outrun: the kicked Solar System destroys itself, of internal causes, on timescales a hundred times shorter than the expected wait for the first disruptive flyby.

\begin{figure*}[!tp]\centering
\includegraphics[width=\textwidth]{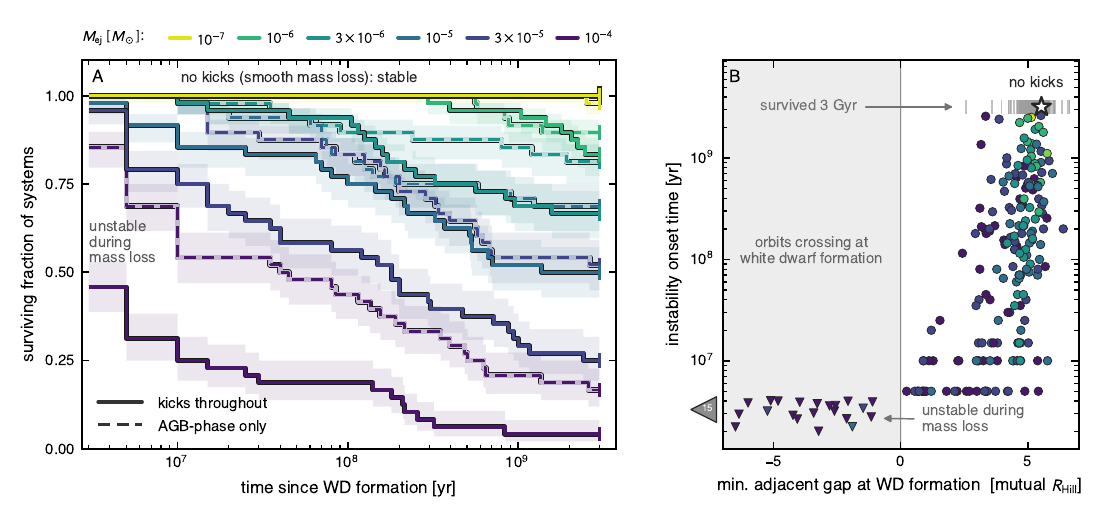}
\caption{\textbf{Survival of the outer Solar System under stochastic mass loss.} Panel A shows the fraction of systems remaining stable versus time since white dwarf formation, per $\Mej$ (colors; solid: ejections throughout the giant-branch phases; dashed: AGB-only), from 3-Gyr integrations without external perturbers. Each curve follows the 48 realizations of each cell; systems that go unstable during the mass loss itself enter the plot at a correspondingly reduced $t=0$ survival value, and onset times are resolved at the 5-Myr cadence of the stability diagnostics. Shaded bands show the 68\% statistical (Wilson) uncertainty of each curve (48 systems per curve); the $3\times10^{-7}\Msun$ curves, statistically indistinguishable from $10^{-7}\Msun$, are omitted for clarity. Panel B depicts the instability onset time versus the minimum adjacent Hill-unit separation of the end state; negative separations denote pairs already reordered or crossing at white dwarf formation. Downward triangles mark systems destabilized during mass loss itself; the count-bearing triangle beyond the left edge tallies the 15 such systems crossing more deeply than the plotted range, and the star marks the unkicked, adiabatically expanded Solar System.}
\end{figure*}

\subsection{Stochastic resonant capture}

Not every architectural rearrangement is destructive, and the random walk exhibits a subtler side effect: occasional capture into mean-motion resonances, unattainable in the adiabatic limit. Capture becomes a first-passage problem, wherein the walk must enter a resonant domain slowly enough to be trapped rather than diffusing through. The associated probability is therefore non-monotonic in $\Mej$ (Fig.~5A), peaking where the accumulated dispersion is comparable to the sum of the resonance's distance from the initial period ratio and its width. Two channels operate over the same range of $\Mej$: the Jupiter--Saturn 5:2 (peak capture probability $16.7^{+6.0}_{-4.7}$\%) and the Uranus--Neptune 2:1 ($20.8^{+6.4}_{-5.2}$\%, AGB-only variant), the latter aided by a self-reinforcing geometry in which the kicks pump Neptune's eccentricity and thereby widen the very domain being approached. Captured pairs librate with near-separatrix amplitudes, pinned at the commensurability ($P_{\rm N}/P_{\rm U} = 2.000\pm0.002$).

\begin{figure*}[!tp]\centering
\includegraphics[width=\textwidth]{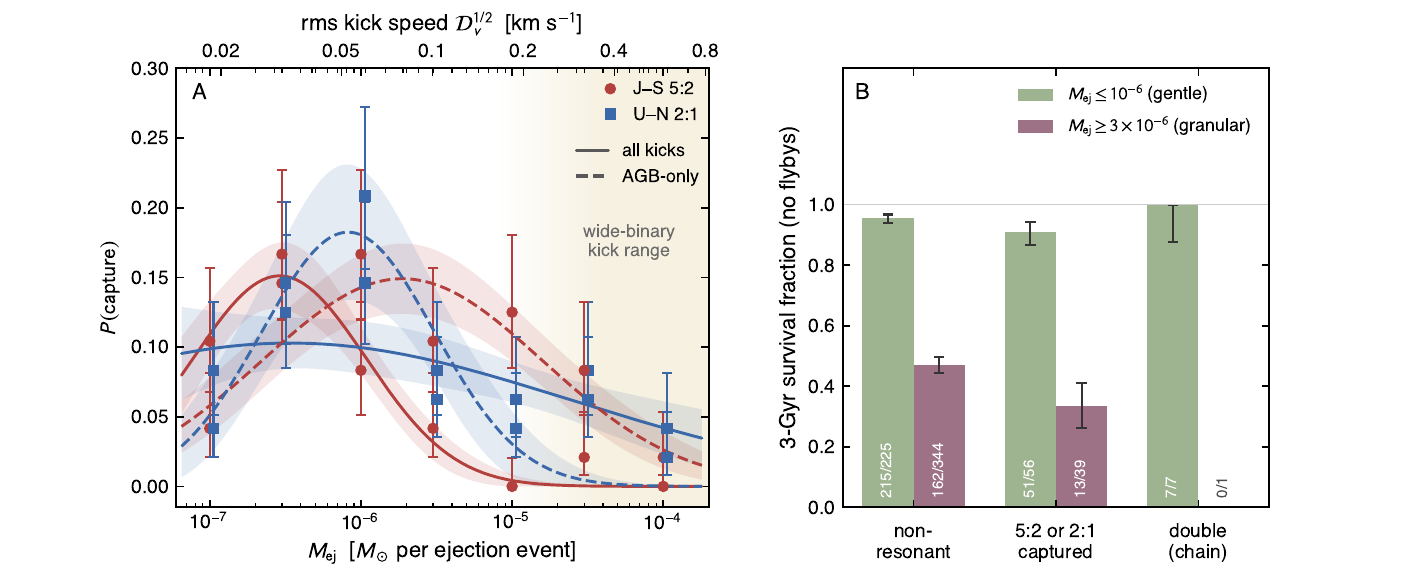}
\caption{\textbf{Stochastic resonant capture and its consequences.} Panel A: Capture probability into the Jupiter--Saturn 5:2 and Uranus--Neptune 2:1 resonances versus $\Mej$ (Wilson 68\% intervals; both variants); the upper axis converts $\Mej$ to the rms diffusion-equivalent recoil under the independent-event assumption (all-phases schedule), and the shaded region marks ejection masses for which that recoil falls within the observationally permitted wide-binary range, and the smooth curves are log-normal profiles in $\log\Mej$ fitted to each series to guide the eye, with shaded 1$\sigma$ fit-uncertainty bands. Panel B: Three-Gyr survival by resonant class in the gentle ($\Mej \le 10^{-6}\Msun$) and granular ($\Mej \ge 3\times10^{-6}\Msun$) regimes; double-resonant systems persist in the gentle regime (the sole granular double-resonant system is disrupted, 0/1), while in the granular regime captured systems show a $\sim$1.8$\sigma$ survival deficit relative to non-resonant ones, consistent with near-separatrix deposition.}
\end{figure*}

In eight of 672 realizations both pairs are trapped simultaneously---a 5:2\,$+$\,2:1 double-resonant configuration of a kind the Solar System has never possessed, whose incidence is consistent with independent trapping and which persists for the full 3\,Gyr wherever assembled in the gentle-$\Mej$ regime. Resonant architectures, familiar as fossils of dissipative capture in young systems such as TRAPPIST-1, can evidently also be assembled by noise around a dying star. Capture is, however, no sanctuary. In the granular-$\Mej$ regime, captured systems survive somewhat \emph{less} often than non-resonant ones (13/39 for singly resonant systems, with the sole granular double-resonant system also disrupting---13/40 combined---versus 162/344 for non-resonant systems, a $\sim$1.8$\sigma$ trend; Fig.~5B), as dissipation-free stochastic capture deposits systems near the chaotic boundary of the resonant domain, where they are unstable without any external perturbations.

\section{Discussion}

The stability of the outer Solar System, long expressed as a ladder of ever-longer timescales---$10^{18}$ years intrinsic \citep{murrayholman1999}, $\sim$$10^{2}$\,Gyr with extrinsic perturbations \citep{zink2020}---thus acquires a rung at the bottom that shortens it dramatically, and the length of that rung is set by a single quantity of stellar physics: the coarseness of solar mass loss. Small $\Mej$ ($\lesssim 10^{-7}\Msun$, the lower edge of our grid and well below model expectations) would largely preserve the frozen adiabatic replica; intermediate values ($10^{-7}$--$10^{-6}\Msun$) yield a $\sim$15--20\% lottery for resonant capture within an otherwise mildly excited system; and $\Mej \gtrsim 10^{-5}\Msun$ produces hot architectures that disassemble themselves within gigayears---or, at the fiducial calibration, while the Sun still shines as a giant. Present evidence favors the violent branch. Under the episodic interpretation, even the conservative lower bound of the permitted kick range, $0.25$\,km\,s$^{-1}$, maps to $\Mej \approx 2\times10^{-5}\Msun$ (all phases) or $6\times10^{-5}\Msun$ (AGB-only)---already in the violent regime---while the nominal $0.75$\,km\,s$^{-1}$ corresponds to $\Mej \approx 1.7$--$5\times10^{-4}\Msun$, at or beyond the top of our grid \citep{elbadryrix2018,hwang2025,fuller2026}. Taken at face value, the observed white dwarf kick distribution thus sets the timescale of the Solar System's dissolution.

Three broader consequences follow. First, under the same episodic interpretation the reasoning extends to planetary systems around every star destined to become a white dwarf, which is to say, nearly all of them. While true Jupiter--Saturn analogues are intrinsically rare (giant-planet occurrence is $\sim$10--15\%; \citealt{fulton2021}), ice-giant-class planets appear to be among the modal outcomes of planet formation \citep{zhudong2021}, and susceptibility to kick-driven disruption scales as $\sqrt{a}\,m_{\rm p}^{-1/3}$ at fixed Hill-unit spacing. This scaling follows from two facts: the recoil of the star perturbs every planet's orbit identically regardless of planet mass, and more weakly bound (larger-$a$) orbits are easier to excite, so the imparted eccentricity diffusion grows as $\sqrt{a}$; meanwhile, the eccentricity required to bridge the gap to a neighboring planet at fixed Hill-unit spacing scales with the Hill radius, $\propto (m_{\rm p}/M_\star)^{1/3}$. The lower-mass systems that dominate the census are therefore more fragile than the Solar System's gas giants at any orbital distance.

Planetary systems, moreover, appear to self-organize toward the edge of stability. This principle is insinuated by the Nice model for the outer Solar System \citep{tsiganis2005} as well as by the ``breaking the chains'' scenario for close-in super-Earth systems \citep{izidoro2017,goldbergpetit2026}, and is quantified by the marginal Hill spacing of observed multi-planet systems \citep{puwu2015}. For such architectures, stochastic mass loss supplies a universal final nudge, i.e., marginal stability maintained for gigayears on the main sequence carries no guarantee through a granular death. The dynamical states of planetary systems around white dwarfs thereby become a fossil record of their hosts' terminal mass-loss granularity.

In this context, the existence of a giant planet on a compact orbit around the white dwarf WD 1856+534 \citep{vanderburg2020} is most readily interpreted as a consequence of intrinsic planet--planet scattering, which can deliver a giant planet to its tidal-circularization radius. More broadly, the ensemble diagnostics are monotonic in the ejection mass---period-ratio dispersion, eccentricity excitation, and destruction fraction all increase with $\Mej$---while resonance occupancy peaks at intermediate $\Mej$, so the statistics of white dwarf planetary systems may one day constrain the coarseness of the mass loss from two sides. 

Second, kicked systems shed planets prolifically, beginning within a few Myr of white dwarf formation, contributing an initially warm population of free-floating planets loosely comoving with young white dwarfs \citep{mroz2017,stephan2026}. In this connection, \citet{stephan2026} have recently shown that the observed kick, applied as a single impulse, destabilizes wide multi-planet systems. Our results resolve that impulse into its physical constituents, and it is precisely this resolution---the $\sqrt{\Mej}$ random walk operating \emph{during} the mass loss---that produces what a single kick cannot: instability commencing during the mass loss itself, the resonant capture pathways and their $\Mej$ dependence, and the calibrated mapping from ejection granularity to dynamical outcome. In a controlled comparison, with each realization's accumulated net kick vector ($|\Delta v| = 0.5\pm0.25$\,km\,s$^{-1}$) applied instead as a single impulse at the end of smooth mass loss, the single-kick systems reach 83\% instability within 3\,Gyr versus 92\% for the random walks. Gross instability is thus nearly saturated in either scenario at these amplitudes, but half of the random-walk-case instabilities occur before the white dwarf even forms, and the prompt ($<$10\,Myr) single-kick fraction of 46\% closely matches the $\sim$47\% found by \citet{stephan2026}.

Third, those delayed disruptions supply late-time dynamical agitation with the timing favored by white dwarf pollution statistics, including onsets broadly distributed in $\log t$ across the simulated 3-Gyr cooling-age range (Fig.~\ref{fig:onsets}), while dispensing with the marginally packed primordial architectures that packing-based instability models must assume \citep{debes2002,mustill2014,veras2016}. The kicks manufacture instability without recourse to marginally packed initial conditions, the actual outer Solar System included (whose present-day Hill spacing renders it immune to packing-based instability altogether), and shift the earliest onsets of instability into the mass-loss epoch itself.

The outcome of the ensuing orbit crossings is governed by the Safronov number $\Theta = (v_{\rm esc,p}/v_{\rm orb})^2 \propto (m_{\rm p}/R_{\rm p})\,(a/M_\star)$. For the Solar System's expanded giants $\Theta \simeq 34$--72, and ejections dominate over collisions, as in our integrations, feeding the free-floating-planet population. Because $\Theta$ depends on orbital distance as much as on planet class, the collision--ejection branching ratio anti-correlates with kick susceptibility: close-in systems have $\Theta \lesssim 1$ and would funnel their instabilities into collisions and star-grazing trajectories---the route that delivers material to the white dwarf. Within a given system, the same logic favors collisions for the lowest-mass planets and ejections for the giants, predicting that white dwarf pollution should be sourced preferentially by rocky bodies rather than by ice- or gas-giant material---consistent with the predominantly rocky compositions inferred for polluted white dwarfs \citep{jurayoung2014}.

We close where the subject began. Newton, contemplating the drifting orbits of Jupiter and Saturn, suspected that the planetary order was mortal, and three centuries of celestial mechanics---from the classical invariability theorems to billion-year numerical integrations---have progressively deferred that verdict, most recently to timescales far beyond the age of the Universe. Our results return the Solar System's dissolution to astrophysically familiar territory, and relocate its cause: not the slow seep of chaos, nor the chance encounter with a passing star, but the Sun itself, which in dying does not merely enlarge the planetary system it built---it shakes it, and more often than not, spills it. Newton's envisioned instability is real after all. He was mistaken only about the perpetrator.

\begin{acknowledgments}
We thank the anonymous referee for a careful review of the manuscript. We are grateful to Ian Brunton, Max Goldberg, and Alessandro Morbidelli for insightful discussions. KB is grateful to the David and Lucile Packard Foundation, the Caltech Center for Comparative Planetary Evolution (3CPE), the National Science Foundation (grant number: AST 2408867), and NASA (Emerging Worlds grant number: 80NSSC26K0395) for their generous support. FCA acknowledges support from NSF Grant No.\ 2508843.
\end{acknowledgments}

\appendix

\section{Numerical setup and implementation of stochastic mass loss}
\label{app:setup}

All integrations use \textsc{rebound} \citep{reinliu2012}: the adaptive high-order IAS15 \citep{reinspiegel2015} through the mass-loss epoch and the hybrid TRACE \citep{lu2024} integrator (timestep 2\,yr) for Gyr-scale stability runs, validated against a smooth-mass-loss control that remains stable and frozen for 3\,Gyr. Instability onset times are chaotic and are interpreted only distributionally; a timestep-halving subsample (12 systems re-integrated at $\Delta t = 1$\,yr) reproduces 11 of 12 stability verdicts, with the single flip and shifted onset times reflecting the expected non-convergence of individual chaotic trajectories, and leaves the per-cell instability fractions unchanged within binomial uncertainties. The Sun and the four giant planets are initialized from JPL Horizons J2000 state vectors. (Initializing instead from tabulated mean orbital elements misplaces the system relative to the 5:2 commensurability---overestimating the period-ratio offset by a factor of two---and yields an incorrect Great-Inequality circulation; state vectors avoid this pitfall.) Our configuration reproduces the observed mean-motion ratio ($2.4833$) and yields a Great-Inequality circulation period of 976\,yr, consistent with the $\sim$900--1{,}000-yr modulation of the real pair. Period ratios quoted in this work are de-aliased mean-motion ratios (the frozen post-mass-loss value is 2.481; the instantaneous osculating ratio oscillates about this with the Great-Inequality amplitude). Mass loss follows a MESA \citep{paxton2019} (release r23.05.1) $1\Msun$, $Z=0.02$, non-rotating track with the same wind prescription as \citet{zink2020} (Reimers $\eta=0.8$; Bl\"ocker $\eta=0.7$; final mass $0.5377\Msun$); the track is time-compressed only where the local mass-loss timescale exceeds 5\,Myr (verified negligible by rate scans spanning $10^3$--$10^6$\,yr); the terminal AGB phase, where ejection events concentrate and all resonant captures occur, is uncompressed.

Ejection events are placed uniformly in ejected mass along the track; each applies an instantaneous $\Delta M = \Mej$ and an isotropic recoil $(\Mej/M_\star)\sqrt{2GM_\star/R_\star(t)}$ to the Sun (impulse duration $\sim$1--2\,yr $\ll$ orbital periods; most events occur near the red-giant- and asymptotic-giant-branch tips, where the mass loss concentrates and kicks are $\approx$5--9\,m\,s$^{-1}$; the largest kicks, $\sim$30\,m\,s$^{-1}$, come from the minority of events during the compact phases bracketing the giant branches, where $v_{\rm esc}$ is largest). The observations constrain the accumulated net recoil $|\sum_i \Delta\mathbf{v}_i|$; planetary excitation is governed instead by the diffusion sum $\mathcal{D}_v = \sum_i |\Delta\mathbf{v}_i|^2 = \sum_i (\Mej/M_{\star,i})^2\,v_{{\rm esc},i}^2$, and the two coincide in expectation, $\langle|\sum_i \Delta\mathbf{v}_i|^2\rangle = \mathcal{D}_v$, only for independently directed, zero-mean events---the defining assumption of the episodic-ejection framework \citep{fuller2026}, which our model adopts and parameterizes through $\Mej$. On the MESA event schedule the accumulated speed $\mathcal{D}_v^{1/2}$ at fiducial $\Mej$ is 0.33 (AGB-only) to 0.58 (all phases) km\,s$^{-1}$, consistent with---though below the nominal $0.75$\,km\,s$^{-1}$ of---the Gaia wide-binary calibration, whose permitted range spans $\sim$0.25--4\,km\,s$^{-1}$, with solar-mass progenitors expected to lie toward the lower end of that range ($\sim$0.5\,km\,s$^{-1}$) \citep{elbadryrix2018,hwang2025}. For $\Mej < 10^{-6}\Msun$, sub-orbital-timescale events are combined with exact $\sqrt{n}$ statistics, where $n$ is the number of events per combined impulse. A mass-step-only control (recoils zeroed) reproduces the adiabatic evolution, isolating the recoil as the operative perturbation. The grid comprises seven values of $\Mej$ log-spaced over $10^{-7}$--$10^{-4}\Msun$, $\times$48 seeds, $\times$2 variants (ejections throughout the giant-branch phases, or restricted to $M_\star < 0.75\Msun$, i.e., AGB-only; episodic behavior is best calibrated on the AGB, so the red-giant-branch extension is an assumption whose consequences the two variants bracket), for a total of 672 realizations.

The phase distribution of the forcing is quantified by the event schedule itself: compact evolutionary stages ($R_\star < 30\,R_\odot$, where $v_{\rm esc}$ is largest) contribute 42\% of $\mathcal{D}_v$ in the all-phases variant (34\% of the Jupiter-orbit-weighted variance), and a matched-seed control battery with recoils suppressed during those stages (24 seeds at fiducial $\Mej$) shows that the prompt channel persists without them: $\sigma(P_{\rm S}/P_{\rm J}) = 0.42$ versus 0.52 (all phases) and 0.20 (AGB-only)---consistent with the schedule-predicted suppression---with orbit crossings during mass loss in 15/24 versus 19/24 realizations (a difference that is itself not significant; two-sided Fisher $p = 0.34$). The all-phases enhancement is therefore carried predominantly by the giant-branch events themselves rather than by the minority of high-$v_{\rm esc}$ kicks from compact stages. (The prominence of the compact stages partly reflects the Reimers wind prescription, which likely overestimates mass loss during compact phases relative to the giant tips; the radius-gated control bounds the consequences of this uncertainty.) Model idealizations---monodisperse $\Mej$ (at fixed total mass loss a spectrum of parcel masses diffuses like a monodisperse population of effective granularity $\langle\Mej^2\rangle/\langle\Mej\rangle$, which folds into the calibration), full momentum coupling at $v_{\rm esc}$ (degenerate with $\Mej$), neglect of direct plume--planet interaction (the parcel's differential gravitational impulse over its orbit-crossing transit at $v_{\rm out}\sim30$\,km\,s$^{-1}$ is bounded at $\lesssim$25\% of the recoil contribution)---are absorbed into the order-of-magnitude uncertainty on $\Mej$ that the grid spans.

Two further controls address the geometry of the ejection. First, stochastically timed but spherically symmetric ejection is null: mass-step-only integrations (recoils zeroed) on the full MESA event schedule, with mass decrements applied instantaneously or delayed by the shell-crossing time, reproduce the adiabatic evolution---final period ratios within the measurement floor of the smooth-mass-loss value, no eccentricity excitation---and remain stable for 3\,Gyr. Second, a latitude-biased ejection distribution is bounded empirically: confining every kick to the equatorial plane at fixed per-event momentum (the configuration that maximizes in-plane forcing) enhances the eccentricity dispersions by the expected factor of at most $\sqrt{3/2}$ (measured $\sigma_{e_{\rm N}}$ ratio 1.27) while leaving the period-ratio dispersion and the 3-Gyr instability fraction unchanged (23/24 versus 22/24); polar-biased distributions err in the opposite, stabilizing direction, and momentum-canceling bipolar ejection is excluded by the observed kicks themselves. Solar oblateness, relativistic precession, tides, the terrestrial planets, the Galactic tide, and stellar flybys are omitted; each is irrelevant at $5$--$60$\,AU on the timescales considered.

\section{Classification, validation, and the correction to Zink et al. (2020)}
\label{app:validation}
\setcounter{figure}{0}
\renewcommand{\thefigure}{B\arabic{figure}}

\subsection{Validity of the impulsive approximation}
Individual ejection events are not instantaneous: the recoil is delivered to the star over the plume-launch time ($\sim$1--2\,yr, of order the pulsation period), and each planet continues to feel the ejected parcel's gravity until the ejecta crosses its orbit ($a/v_{\rm out} \approx 1.6$--9\,yr at $v_{\rm out}\sim30$\,km\,s$^{-1}$). A perturbation of duration $\tau$ acting on an orbit of period $P$ has its effect on the orbital elements suppressed by the transfer function of the pulse shape (for our implementation, a train of ten sub-impulses, the Dirichlet kernel $F = \sin(\pi\tau/P)/[10\sin(\pi\tau/10P)]$), so diffusion coefficients scale as $\overline{F^2}$ averaged over the event schedule. We tested this directly: batteries with each recoil spread over $\tau = 3$, 10, and 30\,yr (kick realizations identical to the instantaneous runs), plus a bracket in which every mass decrement is delayed by 10\,yr.

\begin{figure*}[!tp]\centering
\includegraphics[width=0.95\textwidth]{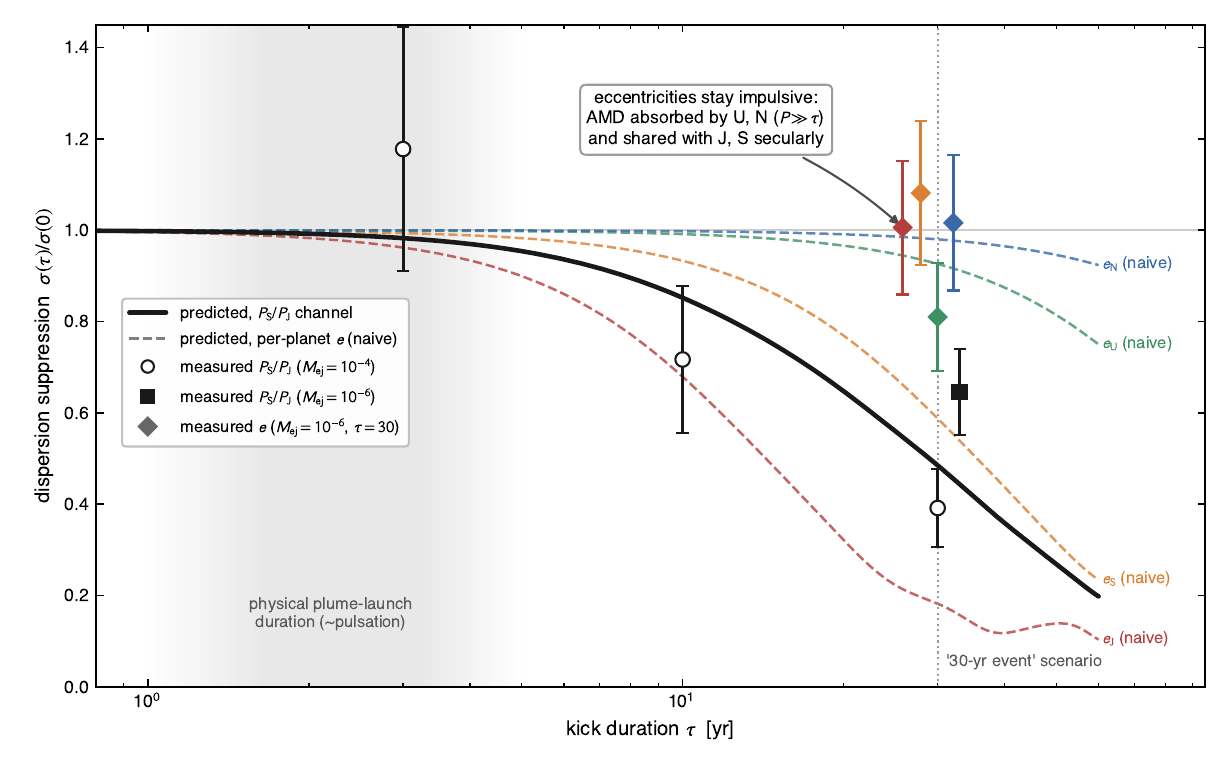}
\caption{\textbf{Validity of the impulsive-kick approximation.} Suppression of kick-driven dispersions when each recoil is spread over a duration $\tau$ (train of ten sub-impulses), relative to instantaneous kicks with identical realizations. The solid curve shows the predicted transfer-function suppression for the Jupiter--Saturn period-ratio random walk, weighted over the ejection-event schedule; measured values (black points; $\Mej = 10^{-4}$ and $10^{-6}\Msun$) follow it. Dashed curves show the corresponding naive per-planet predictions for the eccentricity dispersions; the measured eccentricity dispersions (colored diamonds, $\tau = 30$\,yr; offset horizontally for visibility) instead remain at unity, because angular-momentum-deficit excitation is dominated by Uranus and Neptune---for which any plausible event duration is deeply impulsive ($P \gg \tau$)---and is redistributed to Jupiter and Saturn through secular coupling. Physically motivated event durations (shaded band) lie far inside the impulsive regime for all channels.}
\label{fig:tau}
\end{figure*}

At the physically motivated $\tau \approx 3$\,yr, all dispersions are unchanged within ensemble noise, consistent with the predicted 1--4\% corrections; the mass-delay bracket is likewise null. At an extreme $\tau = 30$\,yr, the one channel to which per-planet transfer theory directly applies---the Jupiter--Saturn period-ratio walk, semimajor axes being immune to secular exchange---shows precisely the predicted suppression ($0.39$ measured versus $0.48$ predicted at fiducial $\Mej$; $0.56$ versus $0.48$ in the diffusive regime after noise-floor correction), validating the transfer-function law at ten times the physical duration. The eccentricity dispersions, by contrast, remain \emph{unsuppressed} even at $\tau=30$\,yr (e.g., $\sigma_{e_{\rm J}}$ ratio $1.01\pm0.15$ against a naive per-planet prediction of 0.18), because the system's angular-momentum deficit is absorbed predominantly by Uranus and Neptune---for which any plausible event duration is deeply impulsive ($P = 84$--570\,yr $\gg \tau$)---and is redistributed to Jupiter and Saturn through secular coupling. Instability statistics are correspondingly unchanged (Fig.~\ref{fig:tau}). More generally, event durations and temporal clustering below the orbital periods renormalize into the effective granularity $\langle\Mej^2\rangle/\langle\Mej\rangle$ per statistically independent impulse; directional correlations between sub-events do not---correlated ejecta must first be summed vectorially into their independent clusters, $\mathcal{D}_v = \sum_a |\sum_{i\in a}\Delta\mathbf{v}_i|^2$, which amounts to promoting the cluster, rather than the parcel, to the role of the independent event.

\subsection{Stability and resonance classification}
Long-term stability is classified by orbit crossing, ejection, or $>$20\% semimajor-axis drift; 3-Gyr verdicts exist for every realization of the grid (672 states); population fractions carry Wilson intervals. Of the 672 verdicts, 448 are stable; the 224 unstable systems first flag as orbit crossing (151), large-scale scattering (49), or ejection (24). These are first-event classifications at the 5-Myr diagnostic cadence, at which several flags can co-occur; the taxonomy is a diagnostic hierarchy rather than a resolved chronology. Following every first-crossing system for a further 100\,Myr beyond its flag resolves the ambiguity at the population level: 113 of the 151 eject at least one planet within the window, 30 proceed to large-scale scattering ($>$20\% semimajor-axis drift), and 8 remain formally crossing without macroscopic evolution; none settles into a phase-protected, non-crossing configuration.

Representative unfoldings are shown in Fig.~3. Post-white dwarf onset times, pooled across $\Mej$ and both variants, span a few Myr to 2.6\,Gyr and are broadly distributed in $\log t$; more than half of the fiducial all-phases realizations are already unstable at white dwarf formation, and the AGB-only median time to instability is $\sim$40\,Myr. During-mass-loss crossings are identified from adjacent-pair perihelion--aphelion overlap in the 10-yr-cadence records. Capture is diagnosed from libration of the critical angles (all four 5:2 components; both 2:1 components) over the final 200\,kyr of each run, using a circular-gap statistic with a winding-number test; capture fractions are per initiated realization, with disrupted systems counted as non-captures. Resonant-domain maps are constructed by direct integration with one planet's semimajor axis scanned across the commensurability.

\begin{figure*}[!tp]\centering
\includegraphics[width=0.9\textwidth]{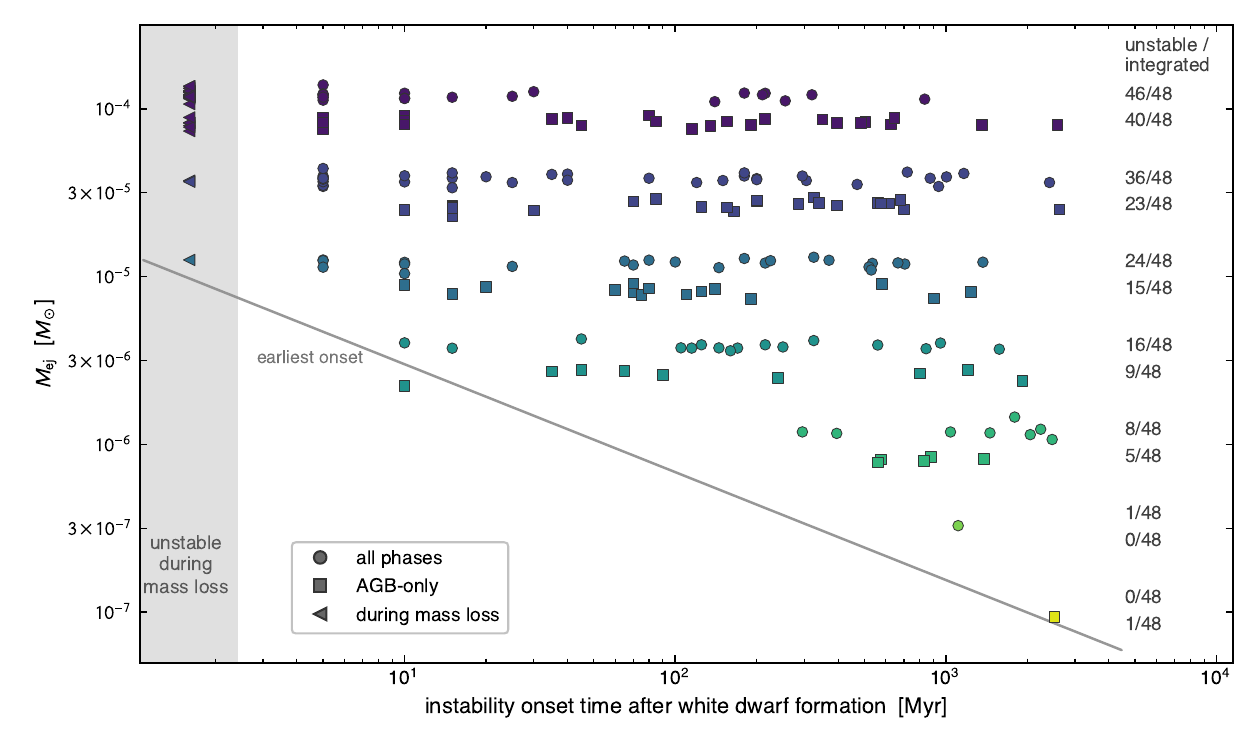}
\caption{\textbf{Instability onset times as a function of ejection granularity.} Each point is one realization's instability onset (circles: ejections throughout the giant-branch phases; squares: AGB-only; leftward triangles: systems already unstable during the mass-loss epoch), from the 48-seed grid of 3-Gyr integrations; fractions at right give unstable/integrated counts. As $\Mej$ grows, the distribution both fills in (higher total instability) and slants toward earlier onsets, terminating at fiducial $\Mej$ in a substantial population that never survives the mass loss at all. The gray curve traces the earliest onset at each $\Mej$.}
\label{fig:onsets}
\end{figure*}

\subsection{The reported 5:2 capture under smooth mass loss}
Three independent lines establish that the adiabatic 5:2 capture reported by \citet{zink2020} is not physical. (i) \emph{Geometry:} the empirically mapped 5:2 domain spans period ratios 2.512--2.535 at $M_\star=0.54\Msun$, entirely above the planets' frozen ratio of 2.481; moreover, as $M_\star$ declines, the domain's lower boundary migrates to \emph{larger} period ratio (from $\simeq$2.506 at $1\Msun$ to $\simeq$2.512 at $0.54\Msun$), so the gap between the resonance and the stationary planets widens as mass loss proceeds. This migration has a transparent origin. The libration domain is not pinned to the Keplerian commensurability, because the resonance condition, $5n_{\rm S} - 2n_{\rm J} - 3\dot\varpi_{\rm S} = 0$, involves the perihelion precession, and both the precession frequencies (in units of the mean motions) and the resonant width are perturbative quantities that scale with the planet-to-star mass ratio. Mass loss leaves the planetary masses untouched while $M_\star$ falls, so the domain's displacement from 2.5 inflates in proportion to $m/M_\star$---the measured offset of the lower boundary indeed doubles (0.006 to 0.012) as $m/M_\star$ grows by the factor 1.85. The planets' period ratio, meanwhile, is frozen by adiabatic invariance; the separatrix recedes from a stationary target. Adiabatic capture is structurally excluded. (ii) \emph{The rate of mass loss:} Zink et al.'s $M_\star(t)$, digitized from the color-coded raster of their Fig.~4 (validated against their quoted masses), peaks at $6\times10^{-6}\Msun\,{\rm yr}^{-1}$---two orders of magnitude below the $\gtrsim 5\times10^{-4}\Msun\,{\rm yr}^{-1}$ we find necessary for non-adiabatic capture---and produces no captures in eight phase realizations integrated through our pipeline (0/8). (iii) \emph{Reproduction of the artifact:} reconstructing the simulation state from orbital elements at a checkpoint while mixing barycentric and heliocentric conventions shifts the period ratio by $+0.2$\% to $+1.5$\% (always toward resonance; exact-protocol controls shift it by $<10^{-5}$), and a subset of such reconstructions capture with $110$--$147^\circ$ libration amplitudes, bracketing the $\sim$110$^\circ$ amplitude measurable in Zink et al.'s figure rasters. Their pipeline contained exactly such a checkpoint (the Phase I/II hand-off at the end of mass loss, coincident with the onset of their reported libration). We therefore regard a coordinate-handling inconsistency at that boundary as the most likely explanation.

\end{document}